\documentclass{elsart}

\usepackage{epsfig}

\usepackage{amssymb}

\begin{document}
\begin{frontmatter}

\title{Two types of $H_{c2}(T)$ dependences in Bi$_2$Sr$_2$Ca$_{1-x}$Y$_x$Cu$_2$O$_{8+\delta}$ with different Yttrium content}

\author[label1,label2]{I. L. Landau}
\author[label1]{H. Keller}
 \address[label1]{Physik-Institut der Universit\"at Z\"urich, Winterthurerstrasse 190, CH-8057 Z\"urich, Switzerland}
 \address[label2]{Institute for Physical Problems, 117334 Moscow, Russia}

%\author{}

%\address{}

\begin{abstract}
	
We reanalyze the magnetization data collected on Bi$_2$Sr$_2$Ca$_{1-x}$Y$_x$Cu$_2$O$_{8+\delta}$ samples (Kim at al, Phys. Rev. B {\bf 72}, 64525 (2005)) and argue that the method, which was used for the analysis of equilibrium magnetization data, is not adequate to the experimental situation. As a result, the temperature dependencies of the upper critical field $H_{c2}(T)$ and the magnetic field penetration depth $\lambda (T)$, obtained in this work, are misinterpreted. Using a different approach to analysis, we demonstrate that the normalized $H_{c2}(T)$ curves are rather different from those presented in the original publication and do not follow predictions of the Werthamer-Helfand-Hohenberg theory. Another important observation is that the $H_{c2}(T)$ dependencies for two samples with different levels of doping are qualitatively different.

\end{abstract}

\end{frontmatter}

%main text

\section{Intrtoduction}

In this paper, we reanalyze mixed-state magnetization $M(H,T)$ data that were collected on several polycrystalline Bi$_2$Sr$_2$Ca$_{1-x}$Y$_x$Cu$_2$O$_{8+\delta}$ (Bi-2212) samples and presented in Ref. \cite{kimcheon}. The main reason is that these data contain some hidden information, which can be discovered by an appropriate analysis. We also argue that the method, which was used for the analysis of equilibrium magnetization data in \cite{kimcheon}, is not adequate to the experimental situation and  the doping dependence of the zero-temperature value of the upper critical field $H_{c2}(0)$, presented in \cite{kimcheon}, should be considered as unjustified.

As was demonstrated in  \cite{lo1,lo-c,lo-tl}, all numerous high-$T_c$ superconductors (HTSC) may be divided into two groups. The dependencies of the normalized upper critical field $h_{c2}$ on $T/T_c$ for HTSC's belonging to the same group are practically identical, while they are distinctly different between the groups. Acknowledging the fact that  the larger group includes a huge variety of different HTSC compounds, while the second one is rather small, we shall denote the corresponding $h_{c2}(T/T_c)$ curves as {\it typical} and {\it unusual}, respectively. Quite surprisingly, the results for two Bi-2212 samples with different levels of doping, investigated in \cite{kimcheon} and analyzed in this work, perfectly match the corresponding $h_{c2}(T/T_c)$ curves for the two above mentioned groups of HTSC's. As well as we are aware, it is the first observation of such a behavior in Bi-based HTSC's. 

There are several theoretical approaches, which are usually employed for evaluation of $H_{c2}$ from experimental magnetization data \cite{hclem,blk,kogan}. All these models assume conventional superconductivity (an isotropic superconducting order parameter) and a uniform sample with a zero demagnetizing factor. Neither of these conditions is satisfied in polycrystalline HTSC's. Because the differences between theoretical assumptions and experimental situations are rarely discussed in the literature, we consider them in some details. 

(i) {\it Demagnetizing factor.} 
If the sample magnetization $M$ is much smaller than an applied magnetic field $H$, demagnetizing effects are usually neglected. This is not correct. If a demagnetizing factor $n \ne 0$, the sample magnetization can be written as $M_{n\ne 0} = (1 - n)M_{n=0}$ (in the case of $4\pi M\ll H$), i.e.,  $M \rightarrow 0$ for $n \rightarrow 1$, independent of an applied magnetic field, temperature or the nature of the sample. Demagnetizing effects are not important if only relative variations of the sample magnetization are considered. In Ref. \cite{kimcheon}, however, the Hao-Clem \cite{hclem} and the vortex fluctuation \cite{blk,kogan} models were employed for the analysis of experimental data. The absolute values of $M$ enter both of these models and neglecting demagnetizing effects may result in misinterpretation of experimental results.

(ii) {\it Pairing symmetry.}
Symmetry of the order parameter is also important. In the case of unconventional $d$-pairing, which is expected in HTSC's, the distribution of the order parameter around vortex cores and the corresponding contribution to the free energy is different from that for conventional superconductors. This is why theoretical calculations based on conventional $s$-pairing should be used  with caution if they are applied for the analysis of experimental data collected on unconventional superconductors

(iii) {\it Polycrystalline samples.}
HTSC's are highly anisotropic. In such materials, if the direction of an external magnetic field does not coincide with one of the principle axes of the crystal, the magnetic induction in the sample is not exactly parallel to the applied magnetic field. In samples consisting of randomly oriented grains, this leads to an additional free energy and may influence the sample magnetization. It should also be noted that, because magnetizations of different grains are different, there is some magnetic interaction between the neighboring grains. The situation is even more complex  at higher temperatures.  Indeed, according to calculations of Brandt \cite{brandt}, the magnetic field dependence of $M$ is a linear function of $\ln H$ (London limit) only in magnetic fields $H < 0.1H_{c2}$ (see also Fig. 3 in Ref. \cite{lo-comm}).  At temperatures, $T \gtrsim 0.8T_c$, the upper limit of the magnetic field range is usually higher than this value. In this case, deviations of $M(H)$ from the predictions of the London model have to be accounted for and a simple averaging, which was proposed in Ref.  \cite{kogan} and used in Ref. \cite{kimcheon}, is not applicable.

The model of thermal fluctuations of vortices \cite{blk}, which was used for the analysis of experimental data in Ref  \cite{kimcheon}, is based on an experimental observation that, in the case of layered HTSC compounds, there is a temperature $T^* < T_c$, at which the sample magnetization does not depend on an applied magnetic field \cite{kes}. As well as we are aware, the model \cite{blk} is the only theory describing this feature. Using this approach, one may evaluate the magnetic field penetration depth $\lambda$ at $T = T^*$. The only parameter $s$, entering the expression for $\lambda(T^*)$, represents the distance between superconducting layers. This parameter may be independently evaluated as $s = - (k_B T^*)/(\Phi _0 M^*)$ ($k_B$ is the Boltzmann constant, and $\Phi _0$ is the magnetic flux quantum). However, the ratio $-T^*/M^*$ is always smaller than the theoretically predicted value and, contrary to the theory, $T^*/M^*$ is practically independent of $s$ \cite{xue}. This is why the model \cite{blk} may be considered only as a qualitative approach to the problem and the resulting value of $\lambda(T^*)$ may be different from the actual magnetic field penetration depth.

One of possible reasons of the above mentioned disagreement between the theory and experiments is that thermal fluctuations of vortices, considered in \cite{blk}, is not the only fluctuation effect that may contribute to the sample magnetization. As was discussed in \cite{lo1}, fluctuations of $T_c$ throughout the sample volume, which cannot be avoided in HTSC's, should also play an important role. Cuprates are non-stoichiometric, which makes them intrinsically inhomogeneous materials. Fluctuations of chemical composition cannot be smaller than the corresponding statistical fluctuations in distribution of non-stoichiometric components. In real samples, however, chemical fluctuations are even stronger than the corresponding statistical numbers. Taking into account that $T_c$ is strongly dependent on the level of oxygen or other dopants, we may conclude that the critical temperature must be spatially dependent.  

At $T$ above the bulk critical temperature, our sample can be considered as superconducting inclusions (grains) imbedded in normal metal. The same situation can be observed at $T < T_c$ in magnetic fields $H > H_{c2}$. At temperatures well below $T_c$ and in $H \ll H_{c2}$, the above mentioned non-uniformity of the sample is not important and it may only lead to weak variations of the vortex density. In this case, the sample magnetization will correspond to an averaged value of $H_{c2}$. However, closer to the $H_{c2}(T)$ line, the situation is different. Indeed, if $H > H_{c2}(T)$, we have only superconducting grains and the diamagnetic moment of the sample should be proportional to $H$ (see Eq. (4.12 in \cite{ovch}). In magnetic fields close but below $H_{c2}(T)$, the superconducting order parameter $\Delta$ is small in the bulk of the sample ($\Delta \rightarrow 0$ for $H \rightarrow H_{c2}$), while it is substantially higher in the regions where local $T_c$ is higher. Considering the sample magnetization, we may conclude that it consists of two diamagnetic contributions: $M_{gr}$ (from inclusions with higher $T_c$)  and $M_{ms}$ (from the mixed state between the inclusions). Both these contributions are approximately linear on $H$.\footnote{According to \cite{brandt}, $M_{ms}(H)$ is approximately linear function down to $H \approx 0.3H_{c2}$.} At the same time, the derivatives $dM_{gr}/dH$ and $dM_{ms}/dH$ have opposite signs. At low temperatures, $M_{ms} \gg M_{gr}$. However, $M_{ms}$ vanishes at the bulk value of $T_c$, while $M_{gr}$ remains non-zero up to somewhat higher temperatures. In other words, it  must be a temperature $T^* < T_c$, at which $(dM_{ms}/dH + dM_{gr}/dH) = 0$, i.e., $M = M_{ms} + M_{gr}$ is temperature independent. 

As we could see, spatial variations of the superconducting critical temperature in inhomogeneous samples result in an effect similar to that of thermal fluctuations of vortices considered in \cite{blk}. It is quite likely that both these effects are important and this is why the interlayer distance $s$, calculated from experimental values of $T^*$ and $M^*$ as $s = - (k_B T^*)/(\Phi _0 M^*)$, is in disagreement with its experimental results. It is also possible that in polycrystalline samples, which are expected to be less homogeneous than single crystals, $T^*$ and $M^*$ are entirely determined by spatial fluctuations. 

\section{Model}

Here, we use a completely different method of analyzing of magnetization data. In this scaling approach, developed in Ref. \cite{lo1},  no particular $M(H)$ dependence is assumed {\it a priori}, and it can be applied to single crystals as well as to polycrystalline samples, independent of the pairing mechanism or the sample geometry \cite{lo-c}. The disadvantage of this analysis is that it does not provide the absolute values of $H_{c2}$ but only its relative temperature variations. This is the prize to pay for its universality.  However, if the value of $H_{c2}$ at any temperature is established, a whole $H_{c2}(T)$ curve is obtained.

The scaling procedure is based on the assumption that the Ginzburg-Landau parameter $\kappa$ is temperature independent. In this case, the mixed-state magnetic susceptibility may be written as
%%%%
\begin{equation}
\chi(H,T) = \chi(H/H_{c2}),
\end{equation}
%%%%
i.e., the temperature dependence of  $\chi$ is only due to temperature variation of $H_{c2}$. Eq. (1) is already sufficient to establish a relation between magnetizations at two different temperatures \cite{lo1} 
%%%%
\begin{equation}
M(H/h_{c2},T_0) = M(H,T)/h_{c2}, 
\end{equation}
%%%%
where $h_{c2}(T) = H_{c2}(T)/H_{c2}(T_0)$ is the upper critical field normalized by its value at some arbitrary chosen temperature $T_0 < T_c$. This equation is valid if the diamagnetic response of the mixed state is the only significant contribution to the sample magnetization. Considering HTSC's, however, we have to take into account their noticeable paramagnetic susceptibility $\chi_n$ in the normal state and its dependence on temperature. In order to account for $\chi_n(T)$, we have to introduce an additional $c_0(T)H$ term in Eq. (2). According to \cite{lo1}, the resulting equation connecting $M(H,T_0)$ and $M(H,T)$ may be written as
%%%%
\begin{equation}
M(H/h_{c2},T_0) = M(H,T)/h_{c2} + c_0(T)H
\end{equation}
%%%%
with
%%%%
\begin{equation}
c_0(T) = \chi_n(T) - \chi_n(T_0)
\end{equation}
%%%%  

We note that Eqs. (1) and (2) are rather general and they can be obtained from any model based on the Ginzburg-Landau theory, including the so-called nonlocal London theory and the Hao-Clem model. At the same time, as was discussed in \cite{lo-hcl,lo-lt2}, these relations remain valid even if $M(H)$ is different from predictions of the conventional Ginzburg-Landau theory and, therefore, they are applicable to unconventional superconductors as well.

It should be noted that at temperatures close to $T_c$, some additional contribution to $M$ arises from fluctuation effects. As was discussed in \cite{lo1}, the second term in Eq. (3) may also account for this contribution. However, because this contribution is not exactly proportional to $H$, it can be accounted for only partially.

Eq. (3) can be used as the basis for the scaling procedure. The adjustable parameters $h_{c2}(T)$ and $c_0(T)$ are obtained from the condition that $M(H/h_{c2},T_0)$, calculated from data collected at different temperatures, collapse onto a single master curve, which represents the equilibrium magnetization at $T = T_0$ \cite{lo1}. As a result, the temperature dependence of the normalized upper critical field $h_{c2}(T)$ is obtained. Extrapolating the resulting $h_{c2}(T)$ curve to $h_{c2} = 0$, we can also obtain the value of the zero-field critical temperature.

It was demonstrated that this scaling procedure works quite well and may be used in order to reliably obtain temperature dependencies of the normalized upper critical field $h_{c2}$ from equilibrium magnetizations measured at different temperatures \cite{lo1,lo-c,lo-comm,lo-hcl,lo-lt2,thomp,doria}. In the following, we use $M_{eff}(H)$ to denote $M(H/h_{c2},T_0)$ calculated using Eq. (3) in order to distinguish it from directly measured magnetization data.

\section{Analysis of magnetization data}

%%%%%%%%%%
\begin{figure}[h]
 \begin{center}
  \epsfxsize=0.8\columnwidth \epsfbox {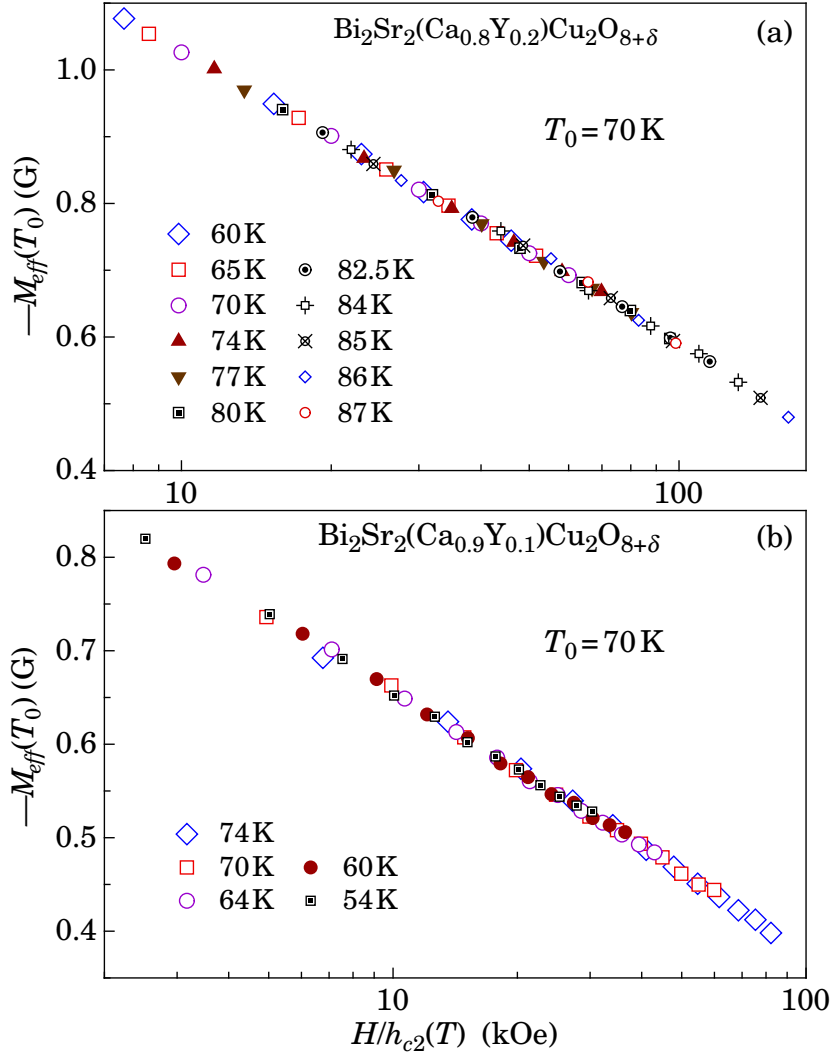}
  \caption{The scaling results for two Bi-2212 samples with different doping levels. Magnetization data are taken from Figs. 1 and 2 of Ref. \cite{kimcheon}.}
 \end{center}
\end{figure}
%%%%%%%%%%
Figs. 1(a) and 1(b) show scaled magnetization curves for two Bi-2212 samples with different Yttrium contents. As may be seen, the quality of scaling is almost perfect in both cases and deviations of individual data points do not exceed the accuracy of the original data as they can be taken from the figures presented in \cite{kimcheon}.

The resulting temperature dependencies of the normalized upper critical field for these samples are shown in Figs. 2 and 3. In order to demonstrate a weak interference between the two fit-parameters, we repeated the scaling procedure assuming $c_0 \equiv 0$. As may be seen in Fig. 2, the difference between the two sets of data-points is insignificant (see also \cite{thomp}). At the same time, because the normal-state paramagnetism in HTSC's exists and it is temperature dependent, we do not see any reason to neglect $c_0(T)$.  

%%%%%%%%%%
\begin{figure}[h]
 \begin{center}
  \epsfxsize=0.8\columnwidth \epsfbox {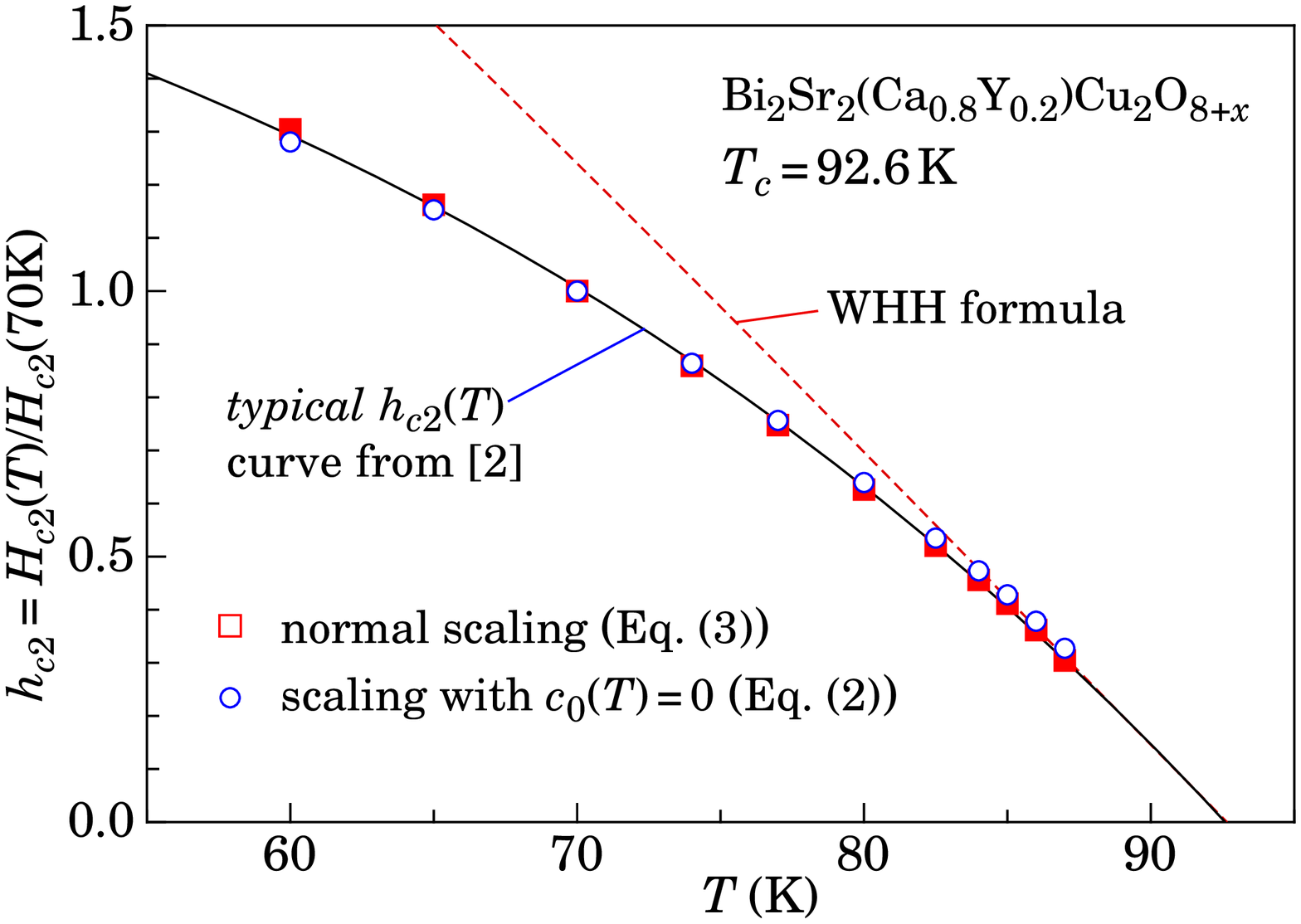}
  \caption{The normalized upper critical field $h_{c2}$ as a function of temperature  for a sample with $x = 0.2$. For comparison, we also show $h_{c2}(T)$ obtained if the scaling is based on Eq. (2) ($c_0(T)  \equiv 0$). The solid line represents the best fit of the "typical" $h_{c2}(T/T_c)$ curve, obtained in Ref. \cite{lo1}, to the data points (see text for details). The dashed line represents to the WHH theory.}
 \end{center}
\end{figure}
%%%%%%%%%%
The results for a sample with the Y-concentration $x = 0.2$ are shown in Fig. 2. The solid line represents the "typical" $h_{c2}(T/T_c)$ dependence, obtained in Ref.  \cite{lo1}, and fitted to the data points by adjusting $T_c$. The value of $T_c = 92.6$ K, evaluated in such a way, is close to the value given in the original publication.\footnote{Because the zero-field superconducting transition in HTSC's is rather broad, the value of $T_c$, evaluated as it was done in \cite{kimcheon}, may be considered only as approximate.} As may be seen in Fig. 2, the data points follow the solid line quite closely clearly demonstrating that this sample belongs to the above mentioned larger group of HTSC's.   We also note that because the $h_{c2}(T)$ curve for this group of HTSC's is quite different from predictions of the Werthamer-Helfand-Hohenberg (WHH) theory \cite{wert} (the dashed line in Fig. 2), the zero-temperature upper critical field $H_{c2}(0)$, evaluated by employing this theory will be well above its real value.  
%%%%%%%%%%
\begin{figure}[h]
 \begin{center}
  \epsfxsize=0.8\columnwidth \epsfbox {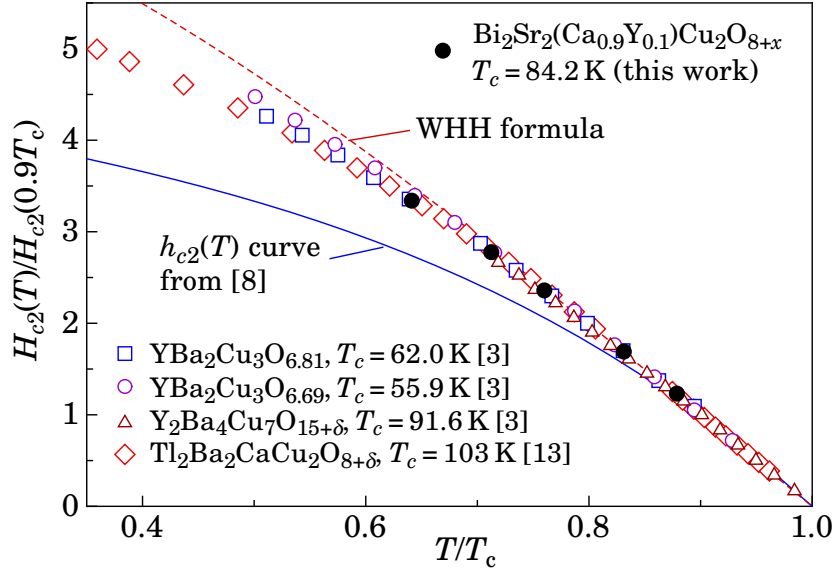}
  \caption{The normalized upper critical field $h_{c2}$ as a function of $T/T_c$  for a sample with $x = 0.1$. The $h_{c2}(T/T_c)$ curves for several samples, belonging to the smaller group of HTSC's \cite{lo-c,lo-hcl}, are shown for comparison. The solid and the dashed lines are the same as in Fig. 2.}
 \end{center}
\end{figure}
%%%%%%%%%%

The $h_{c2}(T/T_c)$ curve for a sample with $x = 0.1$, as shown in Fig. 3, is quite different and practically coincides with those for several other HTSC compounds belonging to the smaller group of HTSC's. Our evaluation of $T_c = 84.2$ K for this sample is practically the same as the value provided in Ref. \cite{kimcheon}.  We note that, although in this case $h_{c2}(T/T_c)$ is closer to the WHH theory, the differences are still significant and this theory should not be used for the evaluation of $H_{c2}(0)$.  

\section{Discussion}

The observation that, considering temperature dependencies of $H_{c2}$, all numerous HTSC's may be divided into two groups is remarkable. However, it is even more surprising that no any intermediate $h_{c2}(T/T_c)$ was observed so far. Because both typical and unusual $h_{c2}(T/T_c)$ curves were observed in the same families of HTSC's \cite{lo1,lo-c,lo-comm}, one may assume that the level of doping is essential. A similar conclusion may also be drown from the results presented in this work. This is why it would be extremely interesting to study the transition from one type of the $h_{c2}(T/T_c)$ dependence to the other systematically. We also note that in this Y-doped Bi-2212,  an overdoped sample fall into the smaller group (unusual $h_{c2}(T/T_c)$ curves), while in the case of Y-123 only underdoped samples behave in such a way (see Fig. 3). 

Finally, we briefly discuss the results of Ref. \cite{kimcheon}. The main result is the dependence of the zero-temperature upper critical field $H_{c2}(0)$ on the doping level. In order to obtained this plot, the authors had to extrapolated their $H_{c2}(T)$ data to $T = 0$. However, as we argue below, neither the $H_{c2}(T)$ curves nor their extrapolations can be considered as reliable.  We also note that the $H_{c2}(T)$ curves presented in \cite{kimcheon} are quite different from our results (see Figs. 2 and 3).

As may be seen in Fig. 4 of Ref. \cite{kimcheon}, the $H_{c2}(T)$ data points cover a rather narrow range of $T/T_c$ values. In such cases, employing theoretical expressions for extrapolation of experimental data is justified only if it is esyablished that the corresponding theory is quantitatively applicable. As well as we are aware, there are no experimental proofs that the modification of the WHH theory, proposed in \cite{bul}, can be used for description of HTSC's. Moreover, there are no theoretical reasons to expect this. Indeed, the theories \cite{wert,bul} are based on the conventional BCS theory and their applicability to unconventional superconductors is questionable. We also note that, considering the $H_{c2}(T)$ data presented in Fig. 4 of Ref. \cite{kimcheon}, one can easily see that the data-points are in disagreement with the theoretical curves, which were used for their extrapolation. 

Another warning is that the presented $H_{c2}(T)$ data do not show a tendency to vanish at $T = T_c$. This is a strong indication on some drawbacks in the theory, which was used for evaluation of $H_{c2}$ from magnetization data. As we discussed in the Introduction, this theory neglects spatial fluctuations of $T_c$ and this is why it is quite likely that it is not applicable to real samples. In this case, the data, presented in Fig.4 of Ref. \cite{kimcheon}, do not represent $H_{c2}$ and the conclusions made in this work are not actually based on the presented experimental data. 

In conclusion, using an alternative approach to the analysis of experimental data presented in Ref. \cite{kimcheon}, we demonstrate that, depending on the Y-content, Bi$_2$Sr$_2$Ca$_{1-x}$Y$_x$Cu$_2$O$_{8+\delta}$ samples may have qualitatively different  temperature dependencies of the upper critical field. This result is in agreement with previous observations of similar behavior in other HTSC compounds \cite{lo-c,lo-hcl}. Another important point is that our $H_{c2}(T)$ curves are rather different from  the results of the original publication. We argue that this disagreement between two analysis of the same experimental data is due to non-applicability of the vortex fluctuation model to description of real HTSC samples.

This work is partly supported by the Swiss National Science Foundation.

\end{document}